\documentclass[letterpaper,twocolumn]{jpsj3}
\usepackage{txfonts}
\usepackage{amssymb, amsmath}
\usepackage{widetext}
\usepackage{here}

\usepackage{graphicx}
\usepackage{dcolumn}
\usepackage{bm}
\usepackage[dvipdfmx]{color}
\usepackage[normalem]{ulem}

\title{Magnetic Switching by Oxygen Adsorption in Metal-Organic Framework Systems}

\author{Masaki Kato$^1$, Kunio Tokushuku$^2$, Hiroyasu Matsuura$^1$, Masafumi Udagawa$^3$ and Masao Ogata$^{1,4}$}
\inst{$^1$Department of Physics, University of Tokyo, Hongo, Bunkyo-ku, Tokyo 113-0033, Japan \\
$^2$Nomura Securities Co., Ltd., Nihonbashi, Chuo-ku, Tokyo 103-8011, Japan \\
$^3$Department of Physics, Gakushuin University, Mejiro, Toshima-ku, Tokyo 171-8588, Japan \\
$^4$Trans-scale Quantum Science Institute, University of Tokyo, Hongo, Bunkyo-ku, Tokyo 113-0033, Japan}

\abst{In this letter, we address magnetization switching by oxygen adsorption in porous metal-organic framework systems. To this end, we construct a simple localized spin model combined with a Langmuir-type formula for oxygen adsorption and study its finite-temperature properties using Monte Carlo simulation. We successfully explain the main features of this phenomenon, such as the discontinuous changes in magnetic states, sensitivity of the magnetic transition temperatures to oxygen pressure, and absence of singularities in adsorbed oxygen. Based on this model, we also reproduce the observed magnetic transition temperatures for a typical value of oxygen adsorption energy.}

\begin{document}
\maketitle

{\it Introduction}.
The Internet of Things (IoT) is now in the spotlight as a basic framework for building next-generation social systems. 
The goal of IoT is to construct systems with intelligent materials that automatically collect the data of their surroundings and mutually exchange information. In this context, sensors play an important role as key devices for constructing such systems. Among the many possible sensors for various stimuli, we focus on a gas sensor, which is required for distinguishing different gas molecules with high accuracy.

To realize such accurate gas sensors, it is desired that the functional materials should react to different gas molecules differently, even if the molecules are similar in size and boiling points. Metal-organic framework (MOF) systems have recently attracted considerable interest as candidates for such materials because of their high designability \cite{kondo1997three,li1998establishing}. Among MOF systems, we focus on porous MOF systems called magnetic sponges ~\cite{B501600M,C0CS00167H,Maspoch:2003aa,ReversibleMagnetism,Ohkoshi:2004aa,Larionova:1997aa,zhang2021metal}. Magnetic sponge materials sensitively change their magnetic states depending on the species of adsorbed gas molecules. This magnetic sensitivity makes it possible to distinguish between different gas molecules based on their magnetic characteristics, even when the molecules have only small differences in their mechanical and thermal properties.

\begin{figure}[h]
\begin{centering}
\includegraphics[width=0.48\textwidth]{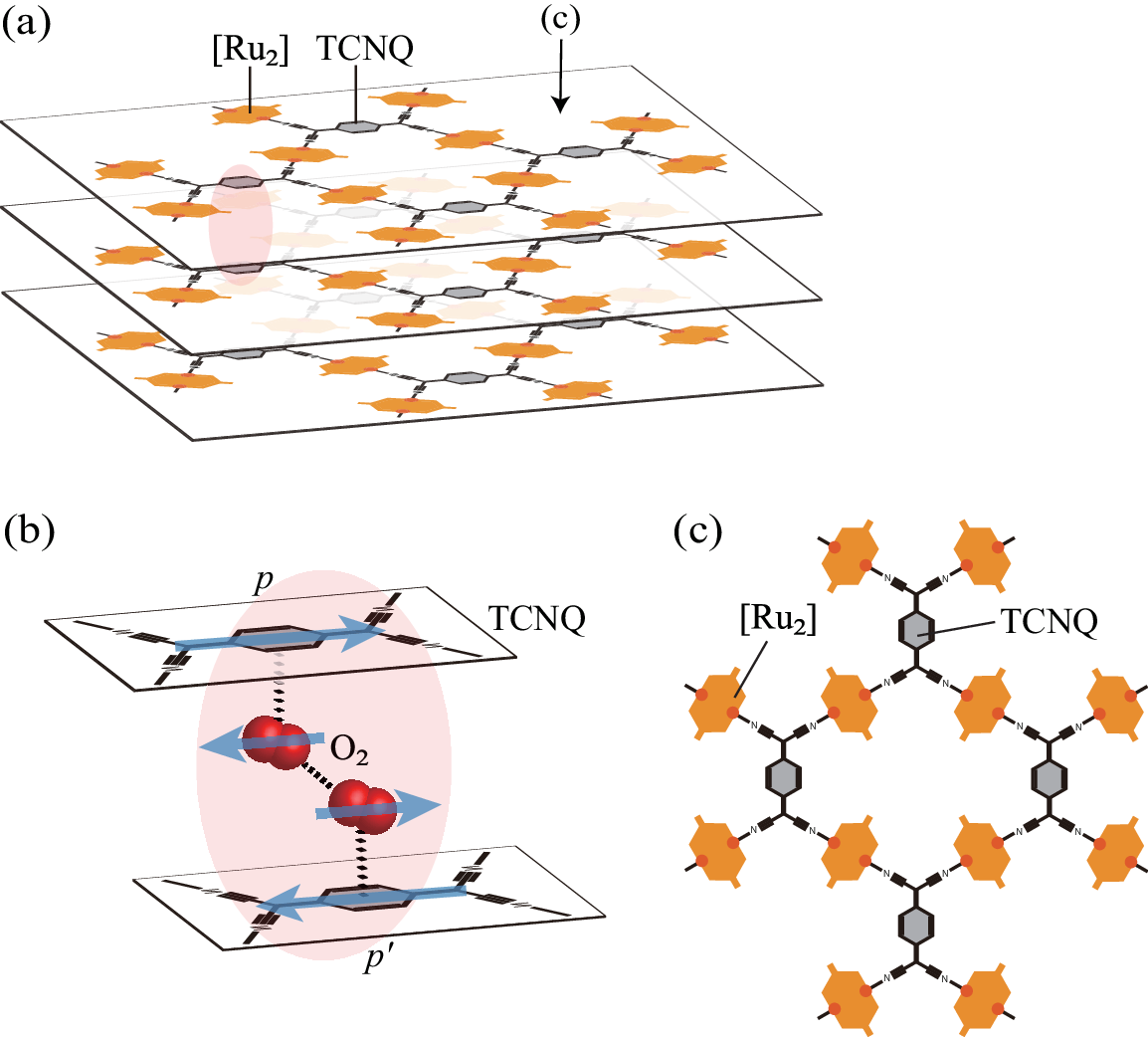}
\caption{(Color online) (a) Schematic diagram of [Ru$_2$]$_2$-TCNQ system. The red-shaded region corresponds to (b). (b) Schematic diagram of the interlayer coupling of TCNQ spins at sites $p$ and $p'$ mediated by adsorbed oxygen molecules. The spins of TCNQ and oxygen molecules are represented by blue arrows. The TCNQ and oxygen spins tend to be antiparallel. The spins of the two oxygen molecules are also aligned antiferromagnetically. The combination of these two antiferromagnetic interactions effectively introduces interlayer antiferromagnetic coupling between the TCNQ spins at sites $p$ and $p'$. (c) Top view of one layer, where [Ru$_2$] and TCNQ spins interact with one another.}
\label{fig:fig1}
\end{centering}
\end{figure}

[\{Ru$_2$(3,5-F$_2$PhCO$_2$)$_4$\}$_2$\{TCNQ(MeO)$_2$\}], which we refer to as [Ru$_2$]$_2$-TCNQ in this letter, is a typical example of a porous MOF system. [Ru$_2$]$_2$-TCNQ is a magnetically active compound with  high gas adsorption capability [Fig.~\ref{fig:fig1}(a)]. This compound has a layered structure of [Ru$_2$]$_2$-TCNQ planes. Because the formal valences of TCNQ(MeO)$_2$ and 3,5-F$_2$PhCO$_2$ are $1-$, the Ru ions take a mixture of [Ru$_2^\mathrm{II, II}$] and [Ru$_2^\mathrm{II, III}$]$^+$ valence states~\cite{kosaka2018gas,kosaka2018layered}. As a result, [Ru$_2$] has localized magnetic moments of $S=1$ or $S=3/2$, and TCNQ has a magnetic moment of $S=1/2$. They interact antiferromagnetically with each other in a two-dimensional network, as shown in Fig.~\ref{fig:fig1}(c)~\cite{Miyasaka:2011aa,Nishio:2014aa}.
In addition to these intralayer magnetic interactions, the magnetic moments are subject to small interlayer magnetic couplings, which are possibly due to dipole interactions~\cite{ReversibleMagnetism,Miyasaka:2006aa,Kosaka:2015aa,kosaka2018layered}.

While the system shows uniform ferrimagnetic ordering below $80\,\rm{K}$ in vacuo, its magnetization profile depends strongly on the gaseous environment. 
When the compound is exposed to gas molecules, such as N$_2$, CO$_2$, and O$_2$, the gas molecules are adsorbed on particular interlayer sites and form chemical bonds connecting the neighboring planes [Fig.~\ref{fig:fig1}(b)]~\cite{kosaka2018gas}. 
In particular, the adsorption of O$_2$ molecules results in a drastic change in the magnetization process~\cite{kosaka2018gas}. In contrast to many other gas molecules, an oxygen molecule has a finite magnetic moment of $S=1$. Therefore, the adsorbed O$_2$ molecules introduce magnetic couplings between the connected TCNQ spins [Fig.~\ref{fig:fig1}(b)] and alter the low-temperature magnetic state of [Ru$_2$]$_2$-TCNQ from ferrimagnetic to antiferromagnetic. In Ref.~\citen{kosaka2018gas}, the swift switching of the uniform magnetization by alternating changes in the O$_2$ pressure was demonstrated.

Despite the remarkable experimental demonstration, theoretical understanding of the switching phenomena is still in its early stages. 
For the [Ru$_2$]$_2$-TCNQ system, there are at least four prominent features to address: 
(i) Magnetic phase transitions: The magnetic state changes drastically as the temperature decreases. [Ru$_2$]$_2$-TCNQ shows a ferrimagnetic transition from the high-temperature paramagnetic phase at $T=T_{\rm F}$. As the temperature is further lowered, a second transition occurs at $T=T_{\rm AF}$ to the antiferromagnetic phase with residual magnetization. (ii) Sensitivity of magnetic states to O$_2$ pressure: The above magnetic transition can be controlled by changing the oxygen pressure, $P$. $T_{\rm AF}$ sensitively changes with $P$, whereas $T_{\rm F}$ is insensitive to the applied pressure. (iii) Absence of singularities in oxygen adsorption: The amount of adsorbed oxygen molecules shows a rather smooth change without singularities at $T=T_{\rm AF}$, despite the sharp change in magnetization at $T=T_{\rm AF}$. (iv) Magnitude of $T_{\rm AF}$: A large discrepancy exists between the magnetic transition temperature $T_{\rm AF}\sim100\,\rm{K}$ and the typical energy scale of oxygen adsorption $\varepsilon_0\sim1000\,\rm{K} \gg T_{\rm AF}$. 

In this work, we construct a minimal statistical mechanical model by which we explain the above features of the O$_2$ adsorption-induced magnetization process of [Ru$_2$]$_2$-TCNQ in a unified manner and establish a general strategy to systematically control the magnetism in gas-adsorbed systems.

{\it Model}.
We start with a Hamiltonian consisting of three degrees of freedom, namely, [Ru$_2$] spins, TCNQ spins, and adsorbed oxygen molecules:
\begin{equation}
\begin{split}
\mathcal{H} = &J_1\sum_{\langle i,p\rangle_{xy}}S_iT_p + J_2\sum_{\langle p,p' \rangle_z}(2\eta n_{pp'}-1)T_pT_{p'} \\
&-\varepsilon_0\sum_{\langle p,p' \rangle_z}n_{pp'}-H\left(g_S\sum_iS_i+g_T\sum_pT_p\right).
\end{split}
\label{eq:Hamiltonian}
\end{equation}
Here, $S_i$ and $T_p$ represent the [Ru$_2$] spin at site $i$ and the TCNQ spin at site $p$, respectively.
We regard them as Ising spins for simplicity and assume that they take binary values; that is, $S_i=\pm 1$ and $T_p=\pm1$. $n_{pp'}$ is an Ising variable representing the O$_2$ adsorption. We consider two cases: $n_{pp'}=1$, if O$_2$ molecules are adsorbed between the TCNQ spins at sites $p$ and $p'$, as shown in Fig.~\ref{fig:fig1}(b), and $n_{pp'}=0$ if no O$_2$ molecules are present. For Zeeman coupling, we introduce $g_T=1/2$ for the TCNQ spins. Because [Ru$_2$] spins are mixtures of $S=1$ and $3/2$, we simply take their average and set $g_S=5/4$ as the effective $g$-factor of [Ru$_2$].

The first term of Eq.~(\ref{eq:Hamiltonian}) represents the intralayer couplings between the [Ru$_2$] and TCNQ spins. These spins are connected to each other, as shown in Fig.~\ref{fig:fig1}(c). We consider only the nearest-neighbor antiferromagnetic coupling $J_1$ as represented by the summation over $\langle i,p\rangle_{xy}$. The second term represents the interlayer couplings between the TCNQ spins. $\langle p,p'\rangle_{z}$ accounts for the summation over pairs on neighboring planes. It is believed that TCNQ spins predominantly couple with one another through long-range magnetic dipolar interactions ~\cite{ReversibleMagnetism,Miyasaka:2006aa,Kosaka:2015aa,kosaka2018layered}. However, in the present study, we consider only the nearest-neighbor components connecting the neighboring layers, which are ferromagnetic, and set their magnitude as $J_2$\cite{Kosaka:2015aa,kosaka2018layered}. We assume $J_2 \ll J_1$. In addition, the adsorption of oxygen molecules leads to a new antiferromagnetic interaction channel, for which the magnitude is set to $2\eta J_2$. 
This emergent coupling between TCNQ spins can be attributed to the magnetic pathway mediated by two oxygen molecules, on which every pair of adjacent spins is aligned antiferromagnetically~\cite{kosaka2018gas} [Fig.~\ref{fig:fig1}(b)].
In the following, we consider the case $\eta>1/2$. The oxygen adsorption thus changes the magnetic coupling from $-J_2(<0)$: ferromagnetic to $(2\eta-1)J_2(>0)$: antiferromagnetic. The third and fourth terms of Eq.~(\ref{eq:Hamiltonian}) represent the bonding energy of the oxygen molecules and the Zeeman energy of the [Ru$_2$] and TCNQ spins, respectively.

The equilibrium state of [Ru$_2$] and TCNQ spins is described by the thermal canonical ensemble defined with the Hamiltonian Eq.~(\ref{eq:Hamiltonian}). Meanwhile, for the oxygen degree of freedom, we assume that the adsorbed oxygen molecules are in equilibrium with oxygen molecules in the gas phase surrounding the system. We describe their equilibrium property using the grand canonical ensemble with the chemical potential $\mu$, which is associated with the oxygen pressure $P$ through the Langmuir formula:
\begin{equation}
\mu=T\ln\left(C\frac{P}{T^{5/2}}\right) \ \ \ \ \left(C= \left(\frac{h^2}{2\pi m_{\mathrm{o}}}\right)^{3/2}\right).
\label{eq:ChemicalPotential}
\end{equation}
Here, $h$ is the Planck constant, and $m_{\mathrm{o}}$ is the mass of an oxygen molecule.
$C\sim4.08\times10^{-7}\,{\rm K}^{5/2}/{\rm kPa}$ is obtained as a combination of these constants. 
Note that the Langmuir formula is based on several assumptions, such as the uniformity of adsorption sites and a limited number of adsorbed molecules, which we assume to be satisfied in our analysis~\cite{butt2013physics}.

In principle, by combining the Hamiltonian in Eq.~(\ref{eq:Hamiltonian}) with the chemical potential of the oxygen molecule, Eq.~(\ref{eq:ChemicalPotential}), we can obtain the full information of our model. However, rather than directly addressing the Hamiltonian, we first trace out $\{S_i\}$ and $n_{pp'}$ and derive an effective Hamiltonian for $\{T_p\}$ of the TCNQ spins. The effective Hamiltonian preserves the information of the original Hamiltonian precisely and is easier to treat numerically. The Ising nature of the [Ru$_2$] spins and oxygen variables enables us to construct the effective model as
\begin{equation}
\mathcal{H}_{\rm eff} = -\frac{1}{\beta}\ln{\rm Tr}_{\{S_i\}, \{n_{pp'}\}}\ e^{-\beta(\mathcal{H}-\mu\sum_{\langle p,p' \rangle_z}n_{pp'})},
\label{eq:EffectiveHamiltonian}
\end{equation}
where $\beta$ denotes the inverse temperature. Up to constant terms, this procedure results in an effective TCNQ Hamiltonian.
\begin{equation}
\mathcal{H}_{\mathrm{eff}} = -J_{\perp}\sum_{\langle p,p'\rangle_z}T_pT_{p'}-J_{\parallel}\sum_{\langle p,p'\rangle_{xy}}T_pT_{p'}-\tilde{H}\sum_p T_p,
\label{eq:EffectiveHamiltonian}
\end{equation}
where the exchange couplings depend on the temperature and chemical potential as a result of tracing out partial degrees of freedom.
\begin{align}
J_{\perp} &= J_2-\frac{1}{2\beta}\ln \left (\frac{1+e^{\beta (\varepsilon_0 + \mu +2\eta J_2)}}{1+e^{\beta (\varepsilon_0 + \mu -2\eta J_2)}} \right),
\label{eq:Jperp}
\end{align}
\begin{align}
J_{\parallel} &= \frac{1}{4\beta}\ln\left (\frac{\cosh(\beta (g_SH+2J_1))\cosh(\beta (g_SH-2J_1))}{\cosh^2(\beta g_SH)} \right)
\end{align}
\begin{align}
\tilde{H} &= g_TH + \frac{1}{\beta}\ln\left(\frac{\cosh(\beta (g_SH-2J_1))}{\cosh(\beta (g_SH+2J_1))} \right).
\end{align}

From this effective model, we can obtain the total magnetization of both the [Ru$_2$] and TCNQ spins per TCNQ molecule as
\begin{equation}
\begin{split}
M_{\mathrm{tot}}  = \frac{1}{N_{\mathrm{TCNQ}}}\left[\right.&g_S \sum_{\langle p,p'\rangle_{xy}}\langle T_pT_{p'}\rangle \\
&+ \left(g_T-2g_S\tanh(2\beta J_1) \right) \sum_p\langle T_p\rangle \left.\right]
\end{split}
\label{eq:TotalMagnetization}
\end{equation}
in the limit of $H\to0$, where $\langle\dots\rangle$ represents the thermal average with respect to $\mathcal{H}_{\mathrm{eff}}$ and $N_{\mathrm{TCNQ}}$ is the number of TCNQ sites. The oxygen density is given by
\begin{equation}
\begin{split}
\langle n_{pp'}\rangle = \frac{1}{2}\left [ \frac{1+\langle T_pT_{p'}\rangle}{e^{\beta (2\eta J_2-\varepsilon_0-\mu)}+1} + \frac{1-\langle T_pT_{p'}\rangle}{e^{\beta (-2\eta J_2-\varepsilon_0-\mu)}+1} \right ].
\end{split}
\label{eq:OxygenDensity}
\end{equation}

We calculate $M_{\rm tot}$ and $\langle n_{pp'}\rangle$ by applying equilibrium Monte Carlo simulations to the effective Hamiltonian, Eq.~(\ref{eq:EffectiveHamiltonian}). The simulation was performed for a system with $50\times50\times20$ sites, which corresponds to 20 layers. At each temperature, we first simulated $10^4$ steps to equilibrate the system and then simulated another $10^5$ steps to measure the physical quantities. Each Monte Carlo step consists of $N_{\rm{TCNQ}}$ single-flip trials. As a realistic parameter set, we chose $J_1=69\,\mathrm{K}$, $J_2=1\,\mathrm{K}$, $\varepsilon_0=1.9\times10^3\,\rm{K}$, and $H=100\,\mathrm{Oe}$~\cite{kosaka2018gas}.
$\varepsilon_0=1.9\times10^3\,\rm{K}$ is of the order of typical values for the physisorption energy\cite{butt2013physics}. Because the actual value of the antiferromagnetic interaction between TCNQ spins is unknown, we used $\eta=2$ as a possible value.
To mimic the experimental cooling of the system, we started with the highest temperature of $140\,\rm{K}$ and repeated the simulation at successively lower temperatures where the final configuration obtained at one temperature was used as the initial state for simulating the next (slightly lower) temperature. Physical quantities were calculated by taking the average over 10 samples with different seeds of random numbers.

\begin{figure}[h]
\begin{centering}
\includegraphics[width = 6.8cm]{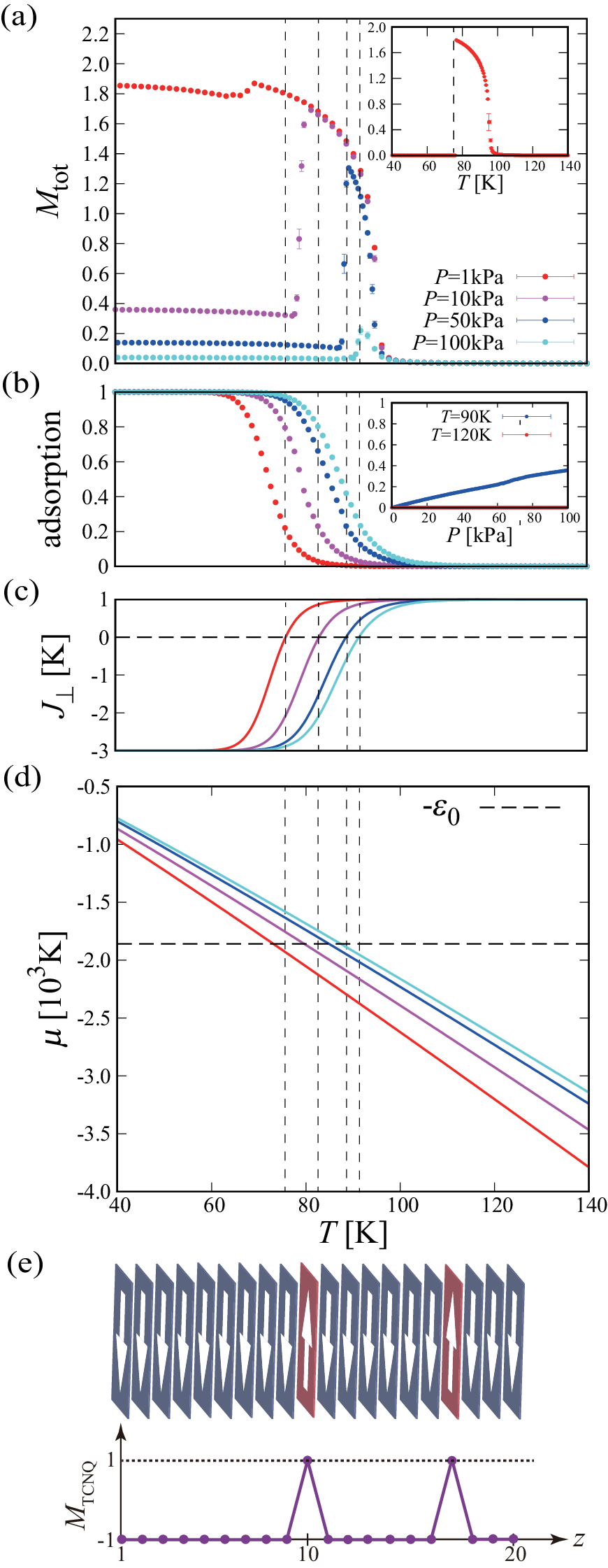}
\caption{(Color online) Temperature dependence of (a) $M_{\mathrm{tot}}$, (b) amount of adsorption, (c) $J_{\perp}$, and (d) $\mu$. (e) (top) Schematic of the magnetization at each layer. (bottom) Average of the TCNQ spins in each layer, $M_{\rm TCNQ}\equiv\langle T_p\rangle_{xy}$. The vertical dashed lines in (a)-(d) indicate the temperature $T_{\rm rev}$ at which $J_1$ changes its sign. The inset of (a) shows the total magnetization at $P=1\,\rm{kPa}$ obtained from the cold-start protocol (see the main text). The inset in (b) shows the $P$ dependence of the oxygen adsorption $\langle n_{pp'}\rangle$.}
\label{fig:fig2}
\end{centering}
\end{figure}

{\it Results}. We present our central results in Fig.~\ref{fig:fig2}. Fig.~\ref{fig:fig2}(a) shows the temperature dependence of the total magnetization $M_{\rm tot}$ for several values of pressure $P$.
This result reproduces the main features of [Ru$_2$]$_2$-TCNQ. The TCNQ spins exhibit ferromagnetic ordering at $T=T_{\rm F}\sim 100\,\rm{K}$ irrespective of the pressure, and a finite magnetization continuously evolves below this temperature, corresponding to the ferrimagnetic ordering observed in the [Ru$_2$]$_2$-TCNQ system. This ferrimagnetic transition can be understood from a mean-field approximation in the two-dimensional plane.

As the temperature is further lowered, the magnetization drops suddenly at $T_{\rm AF}$, which depends on the pressure, implying a first-order phase transition. Below $T_{\rm AF}$, the magnetization takes a finite constant value, with only a slight increase as the temperature decreases, indicating the presence of residual magnetization. A snapshot of the magnetization distribution is shown in Fig.~\ref{fig:fig2}(e). The TCNQ spins are ferromagnetically ordered on one plane, while the ferromagnetic moments alternate in the interlayer direction. Domain formation accounts for the residual magnetization below $T_{\rm AF}$, as discussed later. The entire magnetization process depends sensitively on the applied oxygen pressure $P$. In particular, $T_{\rm AF}$ increases markedly with $P$, whereas $T_{\rm F}$ maintains almost the same value. Eventually, $T_{\rm AF}$ approaches $T_{\rm F}$, and the magnetization exhibits only a tiny kink at $P=100\,\rm{kPa}$. These behaviors are consistent with the experimental results for [Ru$_2$]$_2$-TCNQ, and reproduce features (i) and (ii) in the Introduction.

Fig.~\ref{fig:fig2}(b) shows the amount of adsorbed oxygen molecules, $\langle n_{p,p'}\rangle$, which monotonously grows as the temperature decreases. In particular, $\langle n_{p,p'}\rangle$ shows a rapid increase around $T=T_{\rm AF}$. However, despite the discontinuous change in the magnetization at $T=T_{\rm AF}$, $\langle n_{p,p'}\rangle$ grows smoothly without any apparent singularity around this temperature, reproducing experimental feature (iii) mentioned in the Introduction.

To understand the origin of these behaviors, let us first examine the interlayer coupling, $J_{\perp}$, given in Eq.~(\ref{eq:Jperp}). In our effective model, $J_{\perp}$ depends on the temperature, as shown in Fig.~\ref{fig:fig2}(c). At high temperatures, $J_{\perp}$ takes the bare value of the ferromagnetic coupling $J_2$ between neighboring TCNQ spins. At the lowest temperature $T\sim0\,\rm{K}$, $J_{\perp}$ approaches $(1-2\eta)J_2<0$ because of oxygen adsorption. Accordingly, at an intermediate temperature $T_{\rm rev}$, $J_{\perp}$ vanishes, and sign reversal occurs. From Fig.~\ref{fig:fig2}(c), we find $T_{\rm rev}$ depends on the applied pressure and increases with $P$. In Fig.~\ref{fig:fig2}(a)-(d), $T_{\rm{rev}}$ for each pressure is indicated by a vertical dashed line. Comparing Figs.~\ref{fig:fig2}(a) and (c), it can be observed that $T_{\rm rev}$ is almost equal to $T_{\rm AF}$ at high pressures, while it is slightly higher than $T_{\rm{AF}}$ below $10\,\rm{kPa}$. 
This small discrepancy between $T_{\rm rev}$ and $T_{\rm AF}$ at lower pressures can be attributed to a metastable state. At low pressures, the difference between $T_{\rm rev}$ and $T_{\rm F}$ becomes so large that a large uniform magnetization already develops at $T_{\rm rev}$. Consequently, the system cannot immediately relax to the true ground state. The true ground state of the effective Hamiltonian in Eq.~(\ref{eq:EffectiveHamiltonian}) is the state with perfect inter-plane antiferromagnetic order for which the total magnetization is $M_{\rm{tot}}=0$. In the inset of Fig.~\ref{fig:fig2}(a), we show the magnetization process starting from the true ground state(``cold-start protocol''). In this case, domain wall formation does not occur, and $T_{\rm{AF}}$ coincides with $T_{\rm{rev}}$. The residual magnetization observed in Fig.~\ref{fig:fig2}(a) with decreasing temperature can be attributed to the remaining ferromagnetic domains[Fig.~\ref{fig:fig2}(e)], whose tendency is stronger at lower pressures than at higher pressures.

Because the sign reversal of $J_{\perp}$ is essential in the antiferromagnetic transition, we now focus on the temperature region around $T_{\rm rev}$. 
We first investigate what determines the value of $T_{\rm rev}$. The sign reversal of $J_{\perp}$ results from the competition between the original ferromagnetic interlayer interaction and the oxygen-mediated antiferromagnetic interaction. The latter is controlled by oxygen adsorption. Accordingly, $T_{\rm rev}$ should be well approximated by the characteristic temperature $T_{\rm o}$ at which the chemical potential crosses the adsorption energy $\mu=-\varepsilon_0$. From Eq.~(\ref{eq:ChemicalPotential}), we obtain $T_{\rm o} = \frac{\varepsilon_0}{\alpha}$, where $\alpha\equiv|\ln (CP/T_{\rm o}^{5/2})|$ varies in the range of $25-20$ by changing $P=1-100\,\rm{kPa}$. In the original Langmuir theory, the factor $\alpha$ is related to the entropy density of the oxygen molecule gas $s$ by $\alpha=s-5/2$. In this light, the above discussion suggests that the large energy scale of the bonding energy ($\varepsilon_0\sim 2000\,\rm{K}$) is suppressed to the order of the magnetic transition temperature ($T_{\rm rev}\sim100\,\rm{K}$) by the entropic effect of oxygen molecules; that is, the atmospheric oxygen remains unadsorbed down to a much lower temperature than the bonding energy to maintain a large entropy in the gas phase.
We can therefore explain feature (iv), the large discrepancy between $T_{AF}$ and $\varepsilon_0$. We note that nonmagnetic N$_2$ and CO$_2$ molecules also exhibit steep increases in the adsorption amount at $T=120\,\rm{K}$ and $195\,\rm{K}$, respectively, suggesting similar reduction mechanisms in the adsorption energy scales of these guest molecules. A similar adsorption temperature has also been reported for CO$_2$ adsorption on another magnetic sponge material~\cite{zhang2021metal}.

We note that the small difference between $T_{\rm o}$ and $T_{\rm rev}$ can be accounted for by expanding the Langmuir formula (Eq.~(\ref{eq:ChemicalPotential}) around $T=T_{\rm o}$ as $\mu + \varepsilon_0 \simeq -s(T-T_{\rm o})$ and substituting it into Eq.~(\ref{eq:Jperp}). We then obtain $T_{\rm rev}\simeq(1 + \frac{1}{s}\ln(2\eta-1))T_{\rm o}$ (note that $2\eta J_2 \ll \varepsilon_0$ in Eq.~(\ref{eq:Jperp})). Namely, $T_{\rm rev}$, which is equivalent to the magnetic transition temperature $T_{\rm AF}$, is at most $5-6$\,\% larger than $T_{\rm o}$. 

Finally, we consider the problem of the absence of singularities in $\langle n_{pp'}\rangle$ at $T=T_{\rm AF}$. The oxygen adsorption is obtained from Eq.~(\ref{eq:OxygenDensity}), where $\langle n_{pp'}\rangle$ can be divided into two contributions: $\langle n_{pp'}\rangle=a_+ + a_-\langle T_pT_{p'}\rangle$, where $a_{\pm}=\frac{1}{e^{\beta (2\eta J_2-\varepsilon_0-\mu)}+1} \pm \frac{1}{e^{\beta (-2\eta J_2-\varepsilon_0-\mu)}+1}$. The latter term, which is proportional to $\langle T_pT_{p'}\rangle$, reflects the change in magnetic correlation and is responsible for a possible singularity in $\langle n_{pp'}\rangle$.
However, around $T=T_{\rm AF}\sim T_{\rm rev}$, we have $\frac{J_2}{T}\ll1$ from the difference in energy scales, $T_{\rm rev}\sim100\,\rm{K}$ and $J_2\sim2\,\rm{K}$, and we can approximate $|a_-/a_+|\simeq\tanh\beta2\eta J_2\sim\frac{2\eta J_2}{T}\ll1$. This implies that the oxygen adsorption is mostly determined by the nonmagnetic contribution and is insensitive to the discontinuous change in the magnetic correlation. This explains the absence of a singularity in $\langle n_{pp'}\rangle$.

{\it Summary and Discussion }
In summary, we have studied the magnetization switching of [Ru$_2$]$_2$-TCNQ by constructing a simple spin model combined with the Langmuir formula for oxygen adsorption. In particular, we have explained four main features of this system: (i) the discontinuous changes in the magnetic state, (ii) the sensitivity of the magnetic transition temperatures to oxygen pressure, (iii) the absence of singularities in the adsorbed oxygen, and (iv) the discrepancy between the energy scales of the antiferromagnetic transition and the large bonding energy of oxygen molecules.

The simplicity of our model enables our study to be extended in many directions, including oxygen adsorption beyond the Langmuir model scheme. For example, the presence of several types of adsorption sites and its relation to the second gate opening for oxygen adsorption has been discussed~\cite{kosaka2018gas}.
Perhaps an even more interesting direction is the consideration of possible intralayer magnetic couplings between the TCNQ spins mediated by TCNQ-[Ru$_2$] interactions. The checkerboard-like geometry of Ru ions suggests the possibility of geometrical frustration, which has been intensively studied in two-dimensional systems in various contexts such as macroscopic ground state degeneracy~\cite{Matsuhira_2002,doi:10.1143/JPSJ.72.411,doi:10.1143/JPSJ.73.2845,doi:10.1143/JPSJ.71.2365,PhysRevB.68.064411}, entropic Coulomb interaction~\cite{doi:10.1146/annurev-conmatphys-070909-104138,doi:10.7566/JPSJ.82.073707,takatsu2021universal}, and nontrivial clustering of fractional charges~\cite{PhysRevLett.119.077207,PhysRevB.100.134415,Rau:2016wo}.
Further exploration of [Ru$_2$]$_2$-TCNQ systems may open new avenues in the fertile research field of metal-organic framework materials.

\begin{acknowledgment}
We acknowledge H. Miyasaka for the helpful discussions. 
This work was supported by JSPS KAKENHI (Nos. JP15H05852, JP20H05655, and JP20H04463), MEXT, Japan.
\end{acknowledgment}

\bibliographystyle{jpsjbibstyle.bst}
\bibliography{17785references}

\begin{thebibliography}{10}

\bibitem{kondo1997three}
M.~Kondo, T.~Yoshitomi, H.~Matsuzaka, S.~Kitagawa, and K.~Seki: Angewandte
  Chemie International Edition in English {\bfseries 36} (1997) 1725.

\bibitem{li1998establishing}
H.~Li, M.~Eddaoudi, T.~L. Groy, and O.~Yaghi: Journal of the American Chemical
  Society {\bfseries 120} (1998) 8571.

\bibitem{B501600M}
D.~Maspoch, D.~Ruiz-Molina, and J.~Veciana: Chem. Soc. Rev. {\bfseries 36}
  (2007) 770.

\bibitem{C0CS00167H}
P.~Dechambenoit and J.~R. Long: Chem. Soc. Rev. {\bfseries 40} (2011) 3249.

\bibitem{Maspoch:2003aa}
D.~Maspoch, D.~Ruiz-Molina, K.~Wurst, N.~Domingo, M.~Cavallini, F.~Biscarini,
  J.~Tejada, C.~Rovira, and J.~Veciana: Nature Materials {\bfseries 2} (2003)
  190.

\bibitem{ReversibleMagnetism}
N.~Motokawa, S.~Matsunaga, S.~Takaishi, H.~Miyasaka, M.~Yamashita, and K.~R.
  Dunbar: Journal of the American Chemical Society {\bfseries 132} (2010)
  11943.

\bibitem{Ohkoshi:2004aa}
S.-i. Ohkoshi, K.-i. Arai, Y.~Sato, and K.~Hashimoto: Nature Materials
  {\bfseries 3} (2004) 857.

\bibitem{Larionova:1997aa}
J.~Larionova, S.~A. Chavan, J.~V. Yakhmi, A.~G. Fr{\o}ystein, J.~Sletten,
  C.~Sourisseau, and O.~Kahn: Inorganic Chemistry {\bfseries 36} (1997) 6374.

\bibitem{zhang2021metal}
J.~Zhang, W.~Kosaka, Y.~Kitagawa, and H.~Miyasaka: Nature Chemistry {\bfseries
  13} (2021) 191.

\bibitem{kosaka2018gas}
W.~Kosaka, Z.~Liu, J.~Zhang, Y.~Sato, A.~Hori, R.~Matsuda, S.~Kitagawa, and
  H.~Miyasaka: Nature communications {\bfseries 9} (2018) 1.

\bibitem{kosaka2018layered}
W.~Kosaka, Z.~Liu, and H.~Miyasaka: Dalton Transactions {\bfseries 47} (2018)
  11760.

\bibitem{Miyasaka:2011aa}
H.~Miyasaka, N.~Motokawa, T.~Chiyo, M.~Takemura, M.~Yamashita, H.~Sagayama, and
  T.-h. Arima: Journal of the American Chemical Society {\bfseries 133} (2011)
  5338.

\bibitem{Nishio:2014aa}
M.~Nishio and H.~Miyasaka: Inorg Chem {\bfseries 53} (2014) 4716.

\bibitem{Miyasaka:2006aa}
H.~Miyasaka, T.~Izawa, N.~Takahashi, M.~Yamashita, and K.~R. Dunbar: Journal of
  the American Chemical Society {\bfseries 128} (2006) 11358.

\bibitem{Kosaka:2015aa}
W.~Kosaka, H.~Fukunaga, and H.~Miyasaka: Inorganic Chemistry {\bfseries 54}
  (2015) 10001.

\bibitem{butt2013physics}
H.-J. Butt, K.~Graf, and M.~Kappl: {\em Physics and chemistry of interfaces}
  (John Wiley \& Sons, 2013).

\bibitem{Matsuhira_2002}
K.~Matsuhira, Z.~Hiroi, T.~Tayama, S.~Takagi, and T.~Sakakibara: Journal of
  Physics: Condensed Matter {\bfseries 14} (2002) L559.

\bibitem{doi:10.1143/JPSJ.72.411}
Z.~Hiroi, K.~Matsuhira, S.~Takagi, T.~Tayama, and T.~Sakakibara: Journal of the
  Physical Society of Japan {\bfseries 72} (2003) 411.

\bibitem{doi:10.1143/JPSJ.73.2845}
R.~Higashinaka, H.~Fukazawa, K.~Deguchi, and Y.~Maeno: Journal of the Physical
  Society of Japan {\bfseries 73} (2004) 2845.

\bibitem{doi:10.1143/JPSJ.71.2365}
M.~Udagawa, M.~Ogata, and Z.~Hiroi: Journal of the Physical Society of Japan
  {\bfseries 71} (2002) 2365.

\bibitem{PhysRevB.68.064411}
R.~Moessner and S.~L. Sondhi: Phys. Rev. B {\bfseries 68} (2003) 064411.

\bibitem{doi:10.1146/annurev-conmatphys-070909-104138}
C.~L. Henley: Annual Review of Condensed Matter Physics {\bfseries 1} (2010)
  179.

\bibitem{doi:10.7566/JPSJ.82.073707}
H.~Takatsu, K.~Goto, H.~Otsuka, R.~Higashinaka, K.~Matsubayashi, Y.~Uwatoko,
  and H.~Kadowaki: Journal of the Physical Society of Japan {\bfseries 82}
  (2013) 073707.

\bibitem{takatsu2021universal}
H.~Takatsu, K.~Goto, H.~Otsuka, T.~J. Sato, J.~W. Lynn, K.~Matsubayashi,
  Y.~Uwatoko, R.~Higashinaka, K.~Matsuhira, Z.~Hiroi, et~al.: arXiv:2103.12101
  .

\bibitem{PhysRevLett.119.077207}
T.~Mizoguchi, L.~D.~C. Jaubert, and M.~Udagawa: Phys. Rev. Lett. {\bfseries
  119} (2017) 077207.

\bibitem{PhysRevB.100.134415}
K.~Tokushuku, T.~Mizoguchi, and M.~Udagawa: Phys. Rev. B {\bfseries 100} (2019)
  134415.

\bibitem{Rau:2016wo}
J.~G. Rau and M.~J.~P. Gingras: Nature Communications {\bfseries 7} (2016)
  12234.

\end{thebibliography}

\end{document}